\documentstyle[art10]{article}
\begin{document}

\begin{titlepage}

\title{Octonions: $E_{8}$ Lattice to $\Lambda_{16}$.}

\author{Geoffrey Dixon \\ Department of Mathematics or Physics
\\ Brandeis University \\ Waltham, MA 02254 \\email:
dixon@binah.cc.brandeis.edu
\and Department of Mathematics \\ University of Massachusetts
\\ Boston, MA 02125}

\maketitle

\begin{abstract}
I present here another example of a lattice fibration, a discrete
version of the highest dimensional Hopf fibration: $S^{7}\longrightarrow
S^{15}\longrightarrow S^{8}$.
\end{abstract}

\end{titlepage}

\section*{1. $D_{4}$ to $E_{8}$ with Quaternions.}

To motivate the higher dimensional case involving the octonions, I'll
first develop a lower dimensional lattice fibration using the
quaternions.

Let $q_{m}, \; m=0,1,2,3,$ be a conventional basis for the quaternion
algebra {\bf [1]}, {\bf Q}.  Define
\begin{equation}
D_{4}^{+} = \{\pm q_{m}\} \cup
\{\frac{1}{2}(\pm q_{0}\pm q_{1}\pm q_{2}\pm q_{3})\}.
\end{equation}
These $8+16=24$ unit quaternions form the inner shell
(nearest neighbors to the origin) of a $D_{4} =
\Lambda_{4}$ lattice ($\Lambda_{k}$ is the real laminated lattice in $k$
dimensions {\bf [2]}).  It is well known that the set $D_{4}^{+}$ is
closed under multiplication.

Define
\begin{equation}
D_{4}^{-} = \{\frac{1}{2}(\pm q_{m}\pm q_{n})\}.
\end{equation}
This is also the inner shell of a $D_{4}$ lattice, these
elements normalized to $1/\sqrt{2}$.  This is a (shrunken) $Spin(4)$
rotation of $D_{4}^{+}$.  However,
$D_{4}^{-}$ is not closed under multiplication even if expanded to the
unit sphere.

{}From these two sets we can construct the inner shell of the
8-dimensional  $E_{8} = \Lambda_{8}$  lattice.  In particular,
\begin{equation}
\begin{array}{cll}
& \{<U,0>, \; <0,V>: \; U,V \in D_{4}^{+}\} & (2\times
24 = 48 \mbox{ elements) } \\
\cup & \{<U,V>: \; U,V \in D_{4}^{-}, \; UV^{\dagger}=\pm q_{m}
\} & (8\times 24 = 192 \mbox{ elements)} \\
\end{array}
\end{equation}
($m\in\{0,1,2,3\}$) is the inner shell of an $E_{8}$ lattice, a subset of
the unit 7-sphere in ${\bf Q}^{2}$.

To illustrate the second of the sets in (3), lets look at the set of all
$V \in D_{4}^{-}, \; UV^{\dagger}=\pm q_{m}$ for
$U=\frac{1}{2}(1+q_{1})$.  There are 8 such elements:
\begin{equation}
\begin{array}{lcr}
\pm\frac{1}{2}(1+q_{1}); & \pm\frac{1}{2}(1-q_{1}); &
\frac{1}{2}(\pm q_{2} \pm q_{3}).
\end{array}
\end{equation}
In fact, there are 8 such elements for each $U \in D_{4}^{-}$, hence that
second set has
$8\times 24 = 192$ elements.  The total number of elements is
$48+192 = 240$, which is the order of
$E_{8}$ {\bf [2]}.

There is another characterization of this 192 element subset:
\begin{equation}
\{<U,V>: \; U,V \in D_{4}^{-}, \; V=\pm U \mbox{ or } U+V \in
D_{4}^{+}\}.
\end{equation}
The first two elements in (4) are $\pm U$, and the remaining six
elements satisfy $U+V \in D_{4}^{+}$. \newpage

\section*{2. Fibrations.}

If $U,V \in {\bf Q}$ satisfy
$$
UU^{\dagger}+VV^{\dagger} = 1,
$$
then the doublet $\left[\begin{array}{c} U \\ V \\ \end{array}\right]$
is an element of $S^{7}$, the (unit) 7-sphere.  Define the map
\begin{equation}
\begin{array}{c}
\left[\begin{array}{c} U \\ V \\ \end{array}\right] \longrightarrow
\left[\begin{array}{c} U \\ V \\ \end{array}\right]
\left[\begin{array}{c} U \\ V \\ \end{array}\right]^{\dagger} =
\left[\begin{array}{cc} UU^{\dagger} & UV^{\dagger} \\ VU^{\dagger} &
VV^{\dagger} \\ \end{array}\right] \\ \\
= \frac{1}{2}\left[\begin{array}{cc}1 &0 \\ 0 &1 \\
\end{array}\right] \\ \\
+ \frac{UU^{\dagger}-VV^{\dagger}}{2}\left[\begin{array}{cc}1 &0 \\ 0 &-1
\\ \end{array}\right]
+ \frac{UV^{\dagger}+VU^{\dagger}}{2}\left[\begin{array}{cc}0 &1 \\ 1 &0
\\ \end{array}\right]
+ \frac{UV^{\dagger}-VU^{\dagger}}{2}\left[\begin{array}{cc}0 &1 \\ -1 &0
\\ \end{array}\right].
\end{array}
\end{equation}
The set of all elements
$$
<\frac{UU^{\dagger}-VV^{\dagger}}{2}, \;
\frac{UV^{\dagger}+VU^{\dagger}}{2}, \;
\frac{UV^{\dagger}-VU^{\dagger}}{2}>
$$
(first two real, third pure
quaternion, so 5-dimensional) covers $S^{4}$ (the 4-sphere in ${\bf
R}^{5}$, in this case of radius $\frac{1}{2}$).  The map (6) is an
example of the sphere fibration {\bf [1]}
\begin{equation}
S^{7} \stackrel{S^{3}}{\longrightarrow} S^{4}.
\end{equation}
(Another more interesting example of this fibration in terms of the
octonions was given in {\bf [3]}.)

If $\left[\begin{array}{c} U \\ V \\ \end{array}\right] \in E_{8}$, as
defined in (3), then the map (6) takes $E_{8}$ onto the lattice $Z^{5}$
(all inner shells at this point), consisting of elements of the form
\begin{equation}
\begin{array}{c}
\frac{1}{2}\left[\begin{array}{cc}1 &0 \\ 0 &1 \\
\end{array}\right] \\ \\ \pm
\frac{1}{2}\left[\begin{array}{cc}1 &0 \\ 0 &-1
\\ \end{array}\right] \mbox{ or } \pm
\frac{1}{2}\left[\begin{array}{cc}0 &1 \\ 1 &0
\\ \end{array}\right] \mbox{ or } \pm
\frac{q_{i}}{2}\left[\begin{array}{cc}0 &1 \\ -1 &0
\\ \end{array}\right]
\end{array}
\end{equation}
($i=1,2,3$).  This is an example of the lattice fibration
\begin{equation}
E_{8} \stackrel{D_{4}}{\longrightarrow} Z^{5}.
\end{equation}
(Again, in {\bf [3]} a more interesting example of this fibration was
presented using the octonions.)

\section*{3. $E_{8}$ to $\Lambda_{16}$ with Octonions.}

Let {\bf O} be the octonion algebra {\bf [1,3,4]}.  I choose an octonion
multiplication whose quaternionic triples are determined by the cyclic
product rule,
\begin{equation}
e_{a}e_{a+1} = e_{a+5}, \; a\in\{1,...,7\},
\end{equation}
where the indices in (4) are from 1 to 7, modulo 7 (and
in particular I will set $7=7 \bmod 7$ to avoid confusing $e_{0}$ with
$e_{7}$).  This choice, as it turns out, has an influence on what follows
{\bf [3,4]}.

Define
\begin{equation}
\begin{array}{rl} E_{8}^{+} =  & \{\pm e_{a}\} \\
\cup & \{(\pm e_{a}\pm e_{b}\pm e_{c}\pm e_{d})/2: a,b,c,d
\mbox{ distinct}, \; e_{a}(e_{b}(e_{c}e_{d}))=\pm 1\}, \\ \\
& a,b,c,d\in\{0,...,7\}. \\
\end{array}
\end{equation}
These $16 + 14\times 16 = 240$ elements of the unit
octonion 7-sphere form the inner shell of an
$E_{8}$ lattice, which, like $D_{4}^{+}$ is closed under multiplication
{\bf [5]}.

Define
\begin{equation}
\begin{array}{rl}
E_{8}^{-} =  & \{\frac{1}{2}(\pm e_{a}\pm e_{b}): a,b
\mbox{ distinct}\} \\
\cup & \{\frac{1}{8}(\sum_{a=0}^{7}\pm e_{a}):
\mbox{ odd number of +'s}  \}, \\ \\
\end{array}
\end{equation}  These $112+128 = 240$ elements of the octonion 7-sphere
of radius $1/\sqrt{2}$ also form the inner shell of an
$E_{8}$ lattice, which, like $D_{4}^{-}$ is not closed under
multiplication.  (The effect of my choice of octonion multiplication in
(10) is in the definition of $ E_{8}^{-}$; there are choices that would
require  "odd number of +'s" to be changed to "even number of +'s" in
(12); this would not change the order of that set, which would still be
128.)

{}From these two sets we can construct the inner shell of the
16-dimensional  $\Lambda_{16}$  lattice.  In particular,
\begin{equation}
\begin{array}{cll}
& \{<U,0>, \; <0,V>: \; U,V \in E_{8}^{+}\} & (2\times
240 = 480 \mbox{ elements) } \\
\cup & \{<U,V>: \; U,V \in E_{8}^{-}, \; UV^{\dagger}=\pm e_{a}
\} & (16\times 240 = 3840 \mbox{ elements)} \\
\end{array}
\end{equation}  ($a\in\{0,...,7\}$) is the inner shell of a
$\Lambda_{16}$ lattice, a subset of the unit 15-sphere in ${\bf O}^{2}$.

As an example, let $U = \frac{1}{2}(1+e_{7})$.  Then the 16 values of $V$
for which $<U,V> \in \Lambda_{16}$ are:
$$
\pm U = \pm \frac{1}{2}(1+e_{7}), \; \; \pm \frac{1}{2}(1-e_{7}), \; \;
 \frac{1}{2}(\pm e_{1}\pm e_{5}), \; \; \frac{1}{2}(\pm e_{2}\pm e_{3}),
\; \; \frac{1}{2}(\pm e_{4}\pm e_{6}),
$$
the last 14 of which satisfy $U+V \in E_{8}^{+}$, by which they may also
be characterized.

As another example, let $U =
\frac{1}{4}(-1+e_{1}+e_{2}+e_{3}+e_{4}+e_{5}+e_{6}+e_{7})\in E_{8}^{-}$.
In this case the 16 appropriate $V$'s are:
$$
\begin{array}{r}
\pm U =
\pm \frac{1}{4}(-1+e_{1}+e_{2}+e_{3}+e_{4}+e_{5}+e_{6}+e_{7}), \\
\pm \frac{1}{4}(-1+e_{1}+e_{2}-e_{3}-e_{4}-e_{5}+e_{6}-e_{7}), \\
\pm \frac{1}{4}(-1-e_{1}+e_{2}+e_{3}-e_{4}-e_{5}-e_{6}+e_{7}), \\
\pm \frac{1}{4}(-1+e_{1}-e_{2}+e_{3}+e_{4}-e_{5}-e_{6}-e_{7}), \\
\pm \frac{1}{4}(-1-e_{1}+e_{2}-e_{3}+e_{4}+e_{5}-e_{6}-e_{7}), \\
\pm \frac{1}{4}(-1-e_{1}-e_{2}+e_{3}-e_{4}+e_{5}+e_{6}-e_{7}), \\
\pm \frac{1}{4}(-1-e_{1}-e_{2}-e_{3}+e_{4}-e_{5}+e_{6}+e_{7}), \\
\pm \frac{1}{4}(-1+e_{1}-e_{2}-e_{3}-e_{4}+e_{5}-e_{6}+e_{7}). \\
\end{array}
$$
Again, the last 14 of these elements may be characterized by $U+V \in
E_{8}^{+}$.

\section*{4. More Fibrations.}

If $U,V \in {\bf O}$ satisfy
$$
UU^{\dagger}+VV^{\dagger} = 1,
$$
then the doublet $\left[\begin{array}{c} U \\ V \\ \end{array}\right]$
is an element of $S^{15}$, the (unit) 15-sphere.  As before define the
map
\begin{equation}
\begin{array}{c}
\left[\begin{array}{c} U \\ V \\ \end{array}\right] \longrightarrow
\left[\begin{array}{c} U \\ V \\ \end{array}\right]
\left[\begin{array}{c} U \\ V \\ \end{array}\right]^{\dagger} =
\left[\begin{array}{cc} UU^{\dagger} & UV^{\dagger} \\ VU^{\dagger} &
VV^{\dagger} \\ \end{array}\right] \\ \\ =
\frac{1}{2}\left[\begin{array}{cc}1 &0 \\ 0 &1 \\
\end{array}\right] \\ \\ +
\frac{UU^{\dagger}-VV^{\dagger}}{2}\left[\begin{array}{cc}1 &0 \\ 0 &-1
\\ \end{array}\right] +
\frac{UV^{\dagger}+VU^{\dagger}}{2}\left[\begin{array}{cc}0 &1 \\ 1 &0
\\ \end{array}\right] +
\frac{UV^{\dagger}-VU^{\dagger}}{2}\left[\begin{array}{cc}0 &1 \\ -1 &0
\\ \end{array}\right].
\end{array}
\end{equation}
The set of all elements
$$
<\frac{UU^{\dagger}-VV^{\dagger}}{2}, \;
\frac{UV^{\dagger}+VU^{\dagger}}{2}, \;
\frac{UV^{\dagger}-VU^{\dagger}}{2}>
$$
(first two real, third pure octonion, so 9-dimensional) covers
$S^{8}$ (the 8-sphere in ${\bf R}^{9}$, in this case of radius
$\frac{1}{2}$).  This is the highest dimensional example of a
sphere fibration {\bf [1]}:
\begin{equation} S^{15} \stackrel{S^{7}}{\longrightarrow} S^{8}.
\end{equation}

If $\left[\begin{array}{c} U \\ V \\ \end{array}\right] \in
\Lambda_{16}$, as defined in (13), then the map (14) takes
$\Lambda_{16}$ onto the lattice $Z^{9}$ (inner shells again),
consisting of elements of the form
\begin{equation}
\begin{array}{c}
\frac{1}{2}\left[\begin{array}{cc}1 &0 \\ 0 &1 \\
\end{array}\right] \\ \\ \pm
\frac{1}{2}\left[\begin{array}{cc}1 &0 \\ 0 &-1
\\ \end{array}\right] \mbox{ or } \pm
\frac{1}{2}\left[\begin{array}{cc}0 &1 \\ 1 &0
\\ \end{array}\right] \mbox{ or } \pm
\frac{e_{a}}{2}\left[\begin{array}{cc}0 &1 \\ -1 &0
\\ \end{array}\right]
\end{array}
\end{equation} ($a=1,...,7$).  This is another example of a lattice
fibration:
\begin{equation}  \Lambda_{16} \stackrel{E_{8}}{\longrightarrow} Z^{9}.
\end{equation}

\section*{5. Conclusion.}

The reader may be curious to know the purpose of this work.  If the
reader discovers that purpose before I do, I would ask the reader to let
me know.  For the nonce, it's just pretty stuff.

\end{document}